\documentclass[journal=jpclcd,manuscript=article]{achemso}

\usepackage[version=3]{mhchem} 
\usepackage{esvect}
\usepackage[dvipsnames]{xcolor}
\usepackage{ulem}
\usepackage{pdfpages}

\newcommand{\fluenceunit}{$\mu$J/cm$^2$}

\author{Yulong~Zheng} 
\affiliation[GT-chem]
{School of Chemistry and Biochemistry, Georgia Institute of Technology, 901 Atlantic Drive, Atlanta GA 30332, United~States}

\author{Esteban~Rojas-Gatjens} 
\affiliation[GT-chem]
{School of Chemistry and Biochemistry, Georgia Institute of Technology, 901 Atlantic Drive, Atlanta GA 30332, United~States}

\author{Myeongyeon~Lee}
\affiliation[Leigh]
{Department of Chemical \& Biomolecular Engineering, Lehigh University, 124 E.\ Morton Street, Bethlehem PA 18015, United~States}

\author{Elsa~Reichmanis}
\affiliation[Leigh]
{Department of Chemical \& Biomolecular Engineering, Lehigh University, 124 E.\ Morton Street, Bethlehem PA 18015, United~States}

\author{Carlos~Silva-Acu\~na}
\email{carlos.silva@umontreal.ca}
\affiliation[UdeM]
{Institut Courtois \& D\'epartement de physique, Universit\'e de Montr\'eal, 1375 Avenue Th\'er\`ese-Lavoie-Roux, Montréal, Qu\'ebec H2V~0B3, Canada}
\alsoaffiliation[GT-chem]
{School of Chemistry and Biochemistry, Georgia Institute of Technology, 901 Atlantic Drive, Atlanta GA 30332, United~States}
\title[2D coherent spectroscopy of a push-pull conjugated polymer]
    {Unveiling Multi-Quantum Excitonic Correlations in Push-Pull Polymer Semiconductors}

\begin{document}
\begin{tocentry}
\includegraphics[width=4.9cm, height=5.2cm]{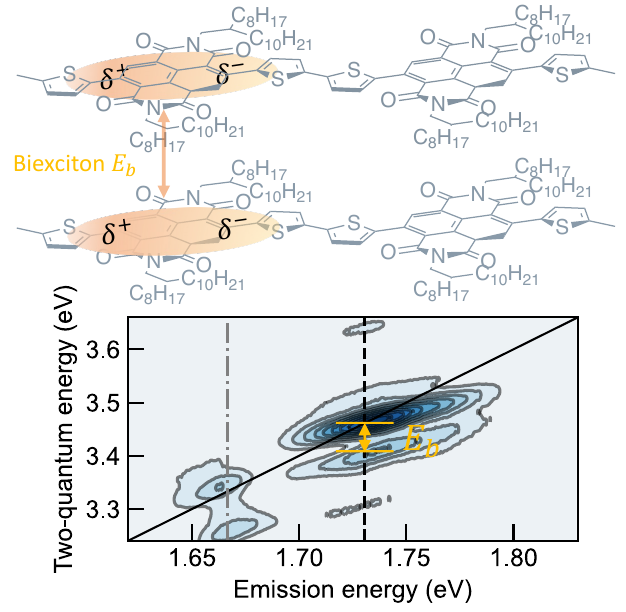}

\end{tocentry}


\begin{abstract}

Bound and unbound Frenkel-exciton pairs are essential transient precursors for a variety of photophysical and biochemical processes. In this work, we identify bound and unbound {Frenkel}-exciton complexes in an electron push-pull polymer semiconductor using coherent two-dimensional spectroscopy. We find that the dominant $A_{0-1}$ peak of the absorption vibronic progression is accompanied by a sub-peak, each dressed by distinct vibrational modes. By considering the Liouville pathways within a two-exciton model, the imbalanced cross peaks in one-quantum rephasing and non-rephasing spectra can be accounted for by the presence of pure biexcitons. The two-quantum non-rephasing spectra, on the other hand, provide direct evidence for unbound exciton pairs and biexcitons with dominantly attractive force. In addition, the spectral features of unbound exciton pairs show mixed absorptive and dispersive character, implying many-body interactions within the correlated {Frenkel}-exciton pairs. Our work offers novel perspectives on the rich photophysical processes in semiconductor polymers with the presence of Frenkel exciton complexes.

\end{abstract}


Frenkel excitons are a collective of local excitations coupled through resonant Coulomb interactions within chromophores.\cite{frenkel1931transformation} Despite the fact that the extent of the delocalization can theoretically span the entirety of the aggregate structure, no realistic molecular aggregates are disorder-free, especially in conjugated polymers, where both static disorder (e.g.\ conformational disorder from site to site) and environmental fluctuations (e.g.\ low-frequency torsional modes) significantly constrain the effective delocalization length.\cite{paquin2013two, spano2014h,moix2013coherent, zheng2023chain} As a consequence, at sufficiently high excitation densities, multiple excitons can coexist in close proximity, leading to distinguishable exciton-exciton interactions and correlations.\cite{knoester2003frenkel} Electron push-pull polymers are known to form disordered polymeric aggregates,\cite{noriega2013general, zheng2017unraveling} which could host two-dimensional hybrid HJ excitons.\cite{yamagata2012interplay, paquin2013two, spano2014h, zhong2020unusual, chang2021hj} In this work, we show direct evidence of two distinct excitons dressed by different vibrational modes, each with its own vibronic progression. Furthermore, we demonstrate the presence of {Frenkel} biexcitons and correlated exciton pairs revealed in one-quantum (1Q) and two-quantum (2Q) two-dimensional coherent spectra (2DCS). By tracing the Liouville pathways qualitatively in a two-exciton basis, we show that the spectral overlap between the biexcitons and the dominant feature of single excitons gives rise to asymmetric cross peaks in the 1Q spectra. 

Under a two-level molecular picture, Frenkel exciton-exciton interactions (EEI) can be categorized into two types: the first type, termed kinematic exciton-exciton interactions, originates from the hard-core-like scattering between Frenkel excitons due to the Pauli exclusion principle.\cite{agranovich2000kinematic} Naturally, the kinematic interaction gives rise to an effective repulsive two-exction state, which is observed as a blue-shifted positive absorption feature (in differential absorption) relative to the ground-state bleach for J-aggregates in pump-probe experiments.\cite{fidder1993observation, bakalis1999pump, jumbo2022cross}. The second type is the dynamic exciton-exciton interaction originating from the differences in the permanent static dipoles of the ground and excited states.\cite{spano1991biexciton} For the latter case, experimental reports on the existence of dynamic Frenkel biexcitons in molecular aggregates are fairly limited, with sporadic evidence provided by fluence-dependent intensity and spectral lineshape analysis by transient absorption measurements.\cite{klimov1998biexcitons, chakrabarti1998evidence, cunningham2020delocalized} Recently, more direct evidence is presented through the use of multi-quantum coherent spectroscopy.\cite{dostal2018direct, maly2020wavelike, maly2020signatures, gutierrez2021frenkel} Dost{\'a}l \textit{et al.} ascribed the growing two-exciton features in a small-molecule aggregated system to exciton-exciton interactions through diffusion.\cite{dostal2018direct} Mal{\'y} \textit{et al.} probed exciton transport transitioning from wavelike to sub-diffusive behavior through EEI in a conjugated copolymer with varying chain lengths.\cite{maly2020wavelike} Guti{\'e}rrez-Meza \textit{et al.} investigated the correlation between the biexcitonic binding energy and hybrid H and J aggregate characteristics in a liquid-crystalline-like conjugated polymer.\cite{gutierrez2021frenkel, bittner2022concerning} Despite the incremental new discoveries of Frenkel-exciton properties, the biexciton resonances and many-body correlations (e.g.\ excitaion-induced dephasing) in Frenkel-exciton systems are not well developed as in their {Wanner-Mott} counterparts.\cite{li2006many,  yang2008revealing, moody2015intrinsic, karaiskaj2010two, thouin2019enhanced, srimath2020stochastic, li2023optical, rojas2023many} In addition, although the biexcitons are observed explicitly in the 2Q spectra, their contributions to the 1Q spectra are often neglected.

\begin{figure}
    \includegraphics[width=0.35\textwidth]{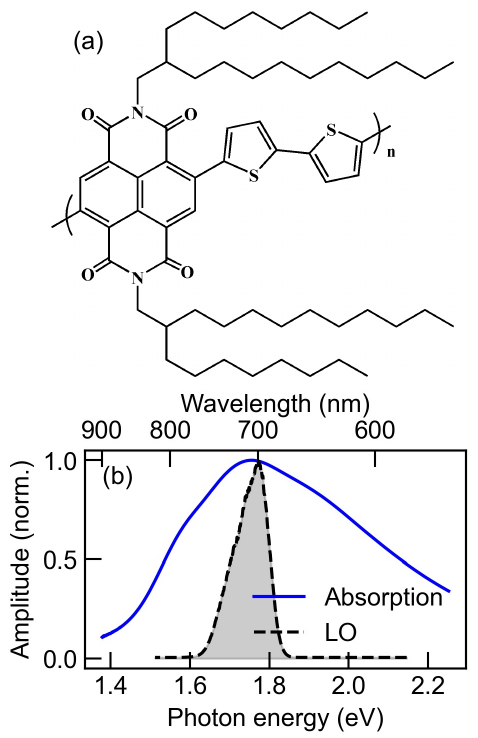}
    \caption{(a) Chemical structure of N2200. (b) Absorption spectra of the low-energy band of N2200 (blue solid line) and the femtosecond pulse spectrum of the local oscillator used in the 2DCS measurements reported in this manuscript (black dashed line shaded in grey).}
    \label{fig1}
\end{figure}

In this {Letter}, we address these issues by probing the conjugated electron push-pull polymer, poly{[N,N'-bis(2-octyldodecyl)naphthalene-1,4,5,8-bis(dicarboximide)-2,6-diyl]-alt-5,5'-\\(2,2'-bithiophene)} (or N2200 and its associated chemical structure is shown in Figure~\ref{fig1}a) by means of two-dimensional coherent spectroscopic measurements. The thin-film preparation of N2200 is described in the Supporting Information (SI). Compared to conjugated homopolymers, the strong charge-transfer character in the electron push-pull polymers leads to a large permanent static dipole moment,\cite{grisanti2009essential} which determines the strength of dynamic EEI.\cite{knoester2003frenkel} Another difference lies in the fact that the electronic transitions in conjugated copolymers are coupled to more vibrational modes, resulting in synergistic intermode effects, where the weakly-coupled mode could borrow intensities from the strongly-coupled vibrational mode.\cite{zhao2007multiple} These, along with polymorphism and static disorder, lead to significantly congested spectral features in electron push-pull polymers.

\begin{figure}
    \centering
    \includegraphics[width=0.5\textwidth]{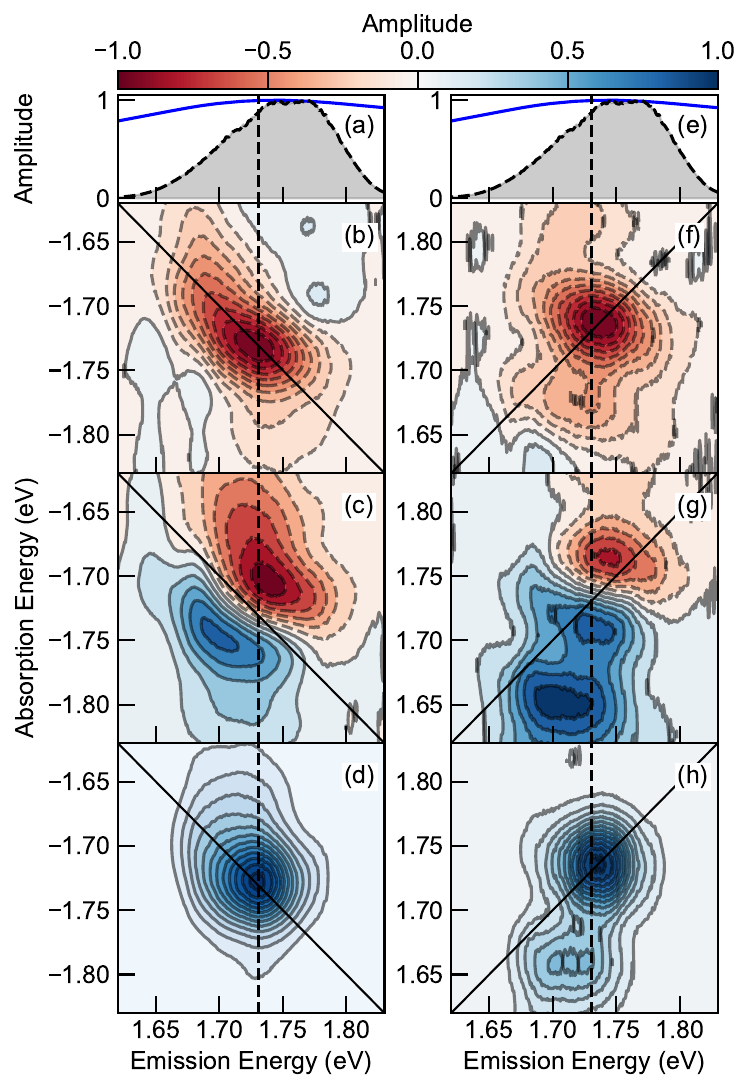}
    \caption{(a) and (e) absorption spectra of $A_{0-1}$ in blue solid curve with the pulse spectra shown in dashed black line shaded in grey. (b)-(d) real, imaginary and absolute spectrum of the rephasing diagram, measured with a fluence of 25.6\,\fluenceunit. (f)-(h) real, imaginary and absolute spectrum of the nonrephasing diagram. All measurements are conducted with the samples positioned in a high-vacuum chamber at ambient temperature.}
    \label{fig2}
\end{figure}

Implementing 2DCS, we directly resolve (i)~electronic correlations between different excited states and (ii)~inhomogeneous and homogeneous broadening contributions into the optical linewidths.\cite{mukamel1995principles} We employ a coherent optical laser beam recombination technique (COLBERT) designed in the research group of Keith Nelson,\cite{vaughan2005diffraction, turner2011invited} which adopts a four-wave-mixing (FWM) signal acquisition scheme based on phase matching imposed by the incident beam geometry and on time ordering of the femtosecond pulse sequence. This spectroscopy generates a third-order macroscopic coherent polarization by interacting a pulse train of three sequential beams with an optically active material resonantly. The coherent emission propagates in the well-defined direction for one-quantum rephasing scheme with wavevector $\vv{k_s}=-\vv{k_a}+\vv{k_b}+\vv{k_c}$ and nonrephasing scheme with wavevector $\vv{k_s}=\vv{k_a}-\vv{k_b}+\vv{k_c}$, where the difference lies on the relative pulse arrival within the two first phase-conjugate pulses. Eventually, the coherent signal is detected through spectral interferometry by an attenuated fourth beam (i.e.\ the local oscillator or LO). Fourier-transforming along the first and third time duration variable gives rise to correlated `absorption' and `emission' axes, where the different electronic transitions lie on the diagonal axis but any correlations between the electronic states show up as cross peaks.\cite{fresch2023two} The experimental method is further explained in SI. In this work, we performed a series of fluence-dependent measurements with the pulse fluence varied from 12.8 to 121\,\fluenceunit at an initial population waiting time of 20\,fs (to avoid contamination from coherent artefacts at shorter delays). Here, we display a case measured at an intermediate fluence in (Figure~\ref{fig2}), with the rest shown in Figure~S2 in SI, where we observed no drastic fluence dependence of the spectral lineshape. We overlap the pulse spectrum with the $A_{0-1}$ vibronic transition (Figure~\ref{fig2}a) in the N2200 thin film. The real, imaginary and absolute part of the 1Q rephasing diagrams are shown in Figure~\ref{fig2}b, c and d, respectively. The real part is absorptive while the imaginary part is dispersive \textit{along} the diagonal axis, as expected in a transmission experiment. Interestingly, the absorptive feature in Figure~\ref{fig2}b appears to be elongated along the diagonal axis, with a slight tail that extends along the absorption energy axis, which is enhanced in the absolute diagram in Figure~\ref{fig2}d. In addition, the cross peaks in Figure~\ref{fig2}b seems to be negative even though they are more attenuated compared to the dominant peaks. We will discuss their implications when considering all possible Liouville pathways later. As the dephasing rate determines the homogeneous linewidth, we also took the antidiagonal cuts of the absolute-valued spectra as shown in Figure~S3. Despite the distinct pumping fluences, we observed no drastic differences between the antidiagonal linewidths, as opposed to what was previously observed in inorganic and perovskite semiconductors, which was attributed to excitation-induced dephasing.\cite{li2006many,moody2015intrinsic,thouin2019enhanced} Such difference might be due to the strongly-bound nature of Frenkel excitons in comparison to Wannier-Mott excitons, where Frenkel excitons are less susceptible to long-range Coulombic screening, at least in this type of push-pull conjugated polymer. In contrast to rephasing spectra, the real (Figure~\ref{fig2}f) and imaginary part (Figure~\ref{fig2}g) of the nonrephasing spectrum demonstrate absorptive and dispersive characteristic \textit{across} the diagonal axis, respectively. A small side peak is observed above the dominant $A_{0-1}$ peak, which might be due to the interstate coherence with $A_{0-2}$ that is out of the spectral range. Since the phase twist is now perpendicular to the diagonal axis, the shoulder at 1.664\,eV along the diagonal axis and cross peak at (1.736, 1.664)\,eV emerge more clearly. The same character is also observed in the absolute-valued diagram for nonrephasing in Figure~\ref{fig2}h.

\begin{figure}
    \centering
    \includegraphics[width=\textwidth]{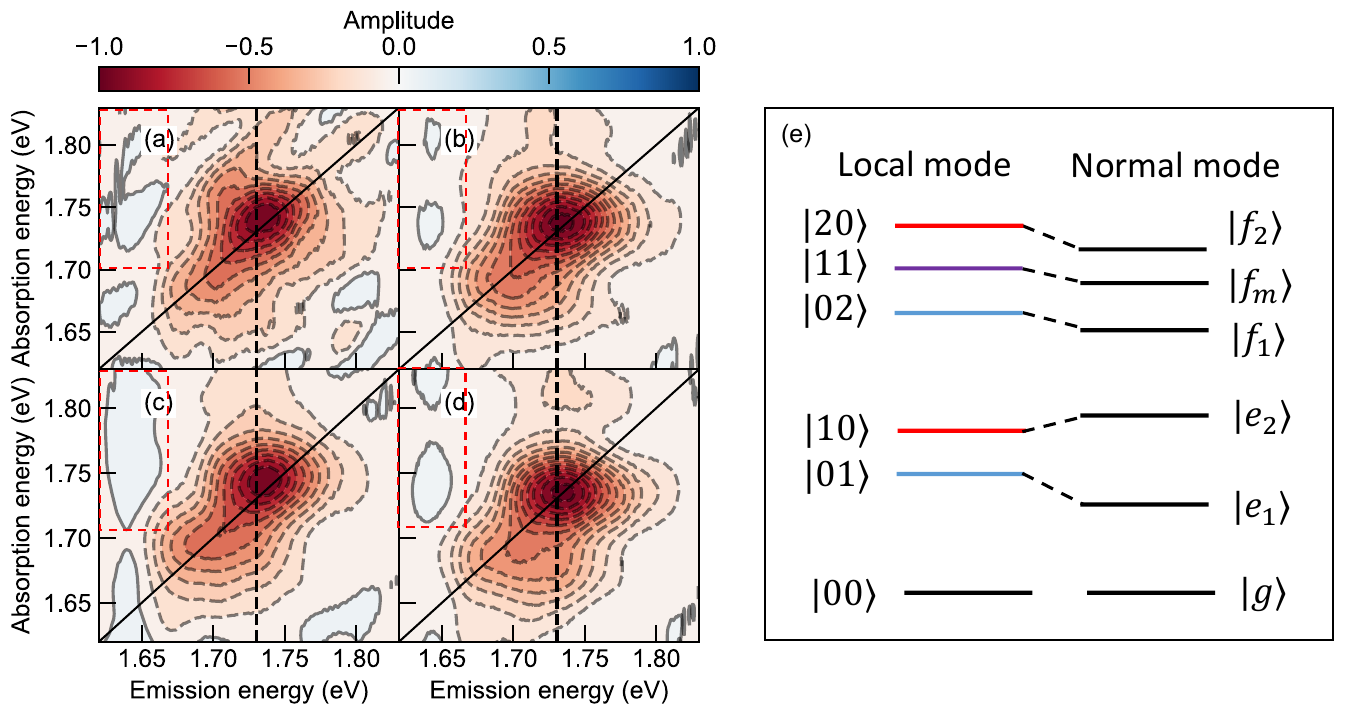}
    \caption{(a)-(d) 1Q total-correlation spectroscopy of 12.8, 25.6, 51.2, and 121\,\fluenceunit, respectively. The black dash lines indicate the position of the dominant $A_{0-1}$ feature. The red dash squares highlight positive features indicating the contribution from biexcitons. (e) Level scheme for both local and normal mode system of two heterogeneous vibronic excitons and their associated biexcitons. The relative energies between the normal modes depend upon the sign and magnitude of exciton-exciton coupling strengths.}
    \label{fig3}
\end{figure}

A common practice to eliminate the phase twist issue is to sum the real part of rephasing and nonrephasing diagram, in which the linewidth is purely absorptive.\cite{hamm2011concepts} By doing so, the small shoulder becomes more prevalent besides the dominant $A_{0-1}$ transition as shown in Figure~\ref{fig3}a-d. It is worth noting that the peak amplitude is weighted by the product of the absorption spectrum and the pulse intensity. As the intensity of the beam is much more attenuated on the low-energy part, this feature should be much stronger than it appears. The cross peak at (1.736, 1.656)\,eV indicates the two excitons share a common ground state, which excludes the possibility of the side peak originating from a different polymer phase. In addition, we also want to highlight that the sub-peak cannot be the tail of $A_{0-0}$ as its energy difference from the $A_{0-1}$ is less than 80\,meV, greatly smaller than the energy of the dominant vinyl-stretching mode (170\,meV), characteristic of various conjugated polymers.\cite{chang2022intermolecular} Another important feature is the asymmetric cross peaks in the upper and lower quadrants of the 2D spectrum, also seen in the nonrephasing diagram. Such signature was explained previously for a cancellation of the Liouville pathways for the interstate coherence and excited state absorption (of mixed biexciton states, see below).\cite{maly2018signatures} The cross peak amplitude would then scale as the coupling strength between the two excitons in the weak-coupling limit. In the upper quadrant, the cross peak shown in the red dashed square has a positive sign as opposed to the dominant features. Since they are overlapped with the dominant feature, their real intensities might be underestimated. 

By identifying the two one-exciton transitions, we can apply the level scheme of a pair of heterogeneous vibronic excitons with their associated biexciton states as shown in Figure~\ref{fig3}e.\cite{hamm2011concepts} $|n_{\nu_1}m_{\nu_2}\rangle$ denotes a state that has $m$ excitons, each coupled to the dominant vinyl-stretching vibrational mode, $\nu_2$, and $n$ excitons coupled to the satellite vibrational mode, $\nu_1$. As the relative position of $|n\rangle$ and $|m\rangle$ can encode the two vibration modes directly, we discard the subscription in the following discussion for simplicity. For FWM experiments, only a conserved two-exciton space needs to be considered. Specifically, we only take account of the pure biexciton states, $|20\rangle$ and $|02\rangle$, and mixed biexciton states, $|11\rangle$. Transitions are ignored when multiple transition dipoles are required. For example, a direct transition, $|10\rangle\rightarrow|02\rangle$ is considered forbidden since it involves multiple photons in one step. The associated normal mode is given schematically on the right panel in Figure~\ref{fig3}e. The exact energy shift depends on the magnitude and sign of the exciton-exciton coupling operators.  

\begin{figure}
    \centering
    \includegraphics[width=\textwidth]{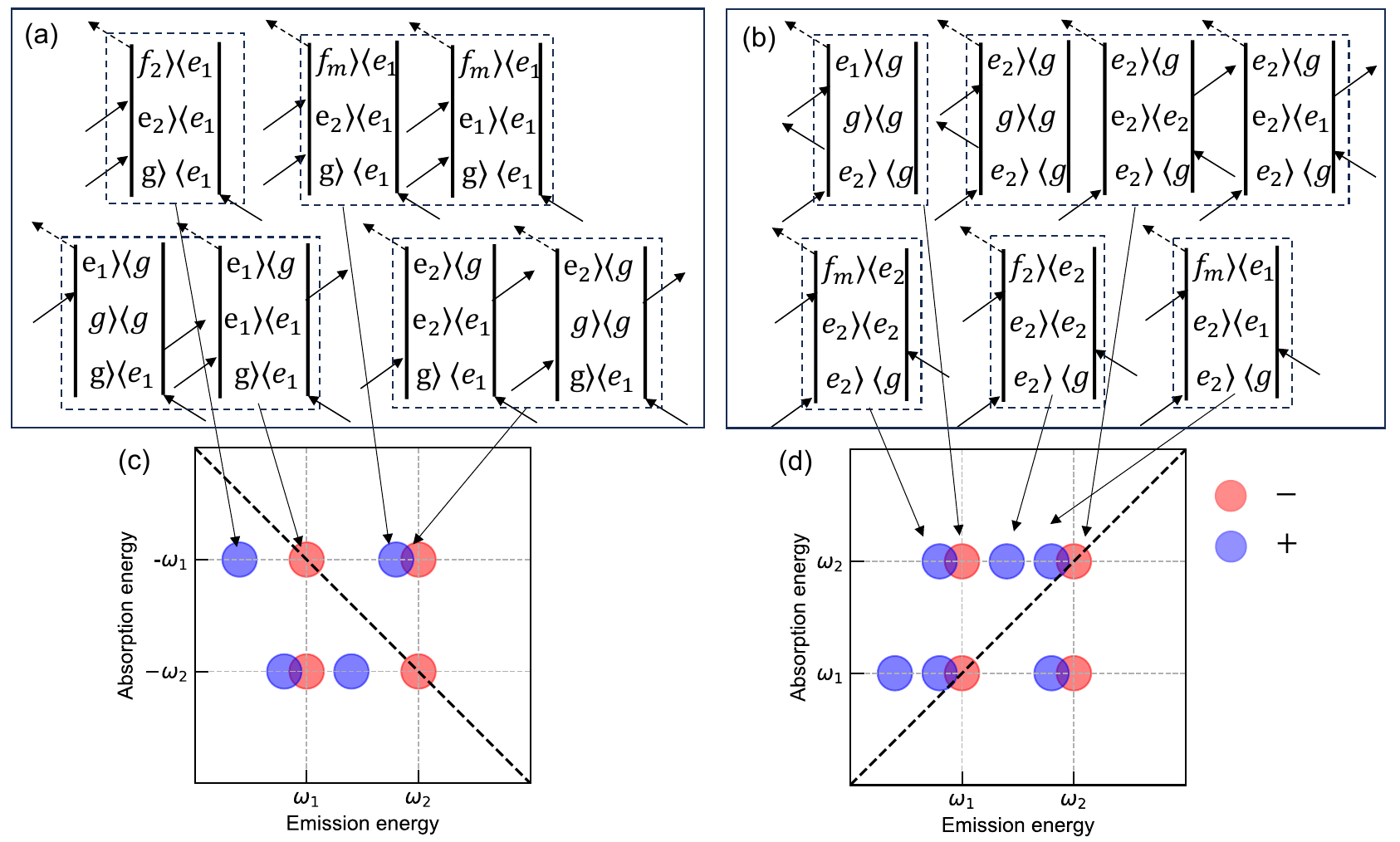}
    \caption{Liouville pathways for rephasing (a) and nonrephasing (b) diagrams. Schematic of purely absorptive 1Q spectra for rephasing (c) and nonrephasing (d) phase matching conditions. The negative and positive features are denoted in red and blue, respectively. Reproduced based on Ref.\citenum{hamm2011concepts}.}
    \label{fig4}
\end{figure}

The Liouville pathways considering all allowed transitions are demonstrated in Figure~\ref{fig4}. Despite the fact that the linewidths in conjugated polymers are broadened, qualitative features can be observed immediately. The positive features that originate from transitions to the mixed biexciton ($|f_m\rangle$) and pure biexciton ($|f_1\rangle, |f_2\rangle$) states, concentrate on the left regime to the diagonal axis, while the right side of the diagonal axis has an overlapped feature from the excited state absorption of the mixed biexciton and interstate coherence as mentioned earlier in both rephasing and nonrephasing spectra.
The overall imbalanced 1Q total correlation spectra are indeed observed as shown in Figure~\ref{fig3}a-d. However, here we considered all three biexciton states, where the pure biexciton features also give rise to positive features even though they could have larger energy shift. Therefore, the overall imbalanced spectral features could be caused by, 1)~the overlap between the mixed biexciton absorption and interstate coherence on both sides of the diagonal axis with their amplitude determined by the EEI strength and their relative transition dipole moments, 2)~pure biexciton absorption on the left regime, leaving the right regime in negative sign due to the interstate coherence solely or 3)~a combination of both contributions. To address this issue, we resort to the 2Q nonrephasing scheme in the phase-matching direction $\vv{k_s}=\vv{k_a}+\vv{k_b}-\vv{k_c}$.\cite{kim2009two} In contrast to the 1Q, now the first two pulses interacting with the material share the same phase. The sequential excitation could generate biexciton and unbound exciton pairs, which can be resolved in more detail along the two-quantum axis.\cite{yang2008two,karaiskaj2010two} 

\begin{figure}
    \centering
    \includegraphics[width=0.8\textwidth]{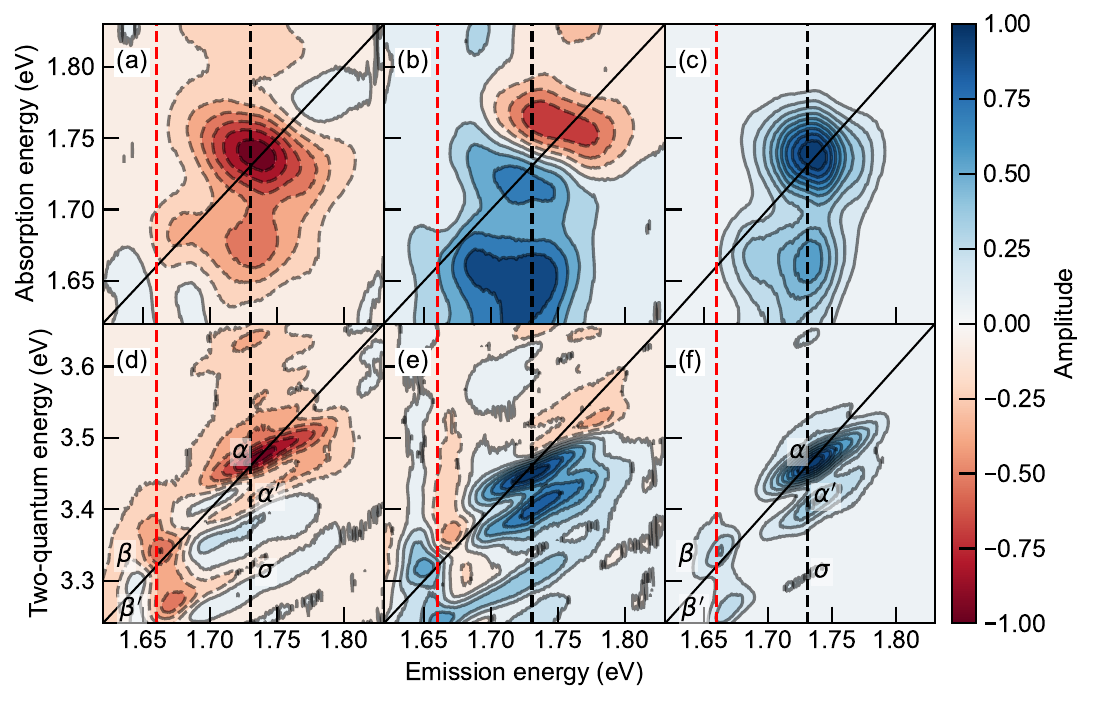}
    \caption{(a) real, (b) imaginary and (c) absolute spectrum of the 1Q nonrephasing spectra. (a) real, (b) imaginary and (c) absolute spectrum of the 2Q nonrephasing diagram. The black dash line indicate the peak position at the dominant $A_{0-1}$. The red dash line locates at the side peak position.}
    \label{fig5}
\end{figure}

To more easily compare the 2Q spectral features with their 1Q counterparts, the 1Q and 2Q nonrephasing measurements are presented in Figure~\ref{fig5} under pumping fluence of each pulse being 121\,\fluenceunit. The spectral features do not significantly depend on the fluences, although low fluence measurements seem to present more artefacts as shown in Figure~S4 in SI. A close match of the energies of the two heterogeneous vibronic excitons in 1Q and 2Q spectra can be found by the red and black dashed lines. Two prominent features can be observed and explained in Figure~\ref{fig5}d-f. First, the two dominant peaks, $\beta$ and $\alpha$ reside on the diagonal axis, each accompanied by a red-shifted side peak, $\beta'$ and $\alpha'$ respectively, along the two-quantum axis. Therefore, the binding energies, experimentally determined as ($E_{2Q}-2E_{1Q}$), for $|f_1\rangle$ and $|f_2\rangle$ are estimated to be -76 and -64\,meV as shown in Figure~S6, respectively, where the negative sign indicates their attractive nature. The exciton binding energies are comparable since the two vibronic excitons have the same electronic origins, while the slight difference might originate from the perturbation of the two distinct vibrational modes. Interestingly, a blue-shifted shoulder around (1.736, 3.502),eV can also be observed extending out of $\alpha$. Therefore, the repulsive binding energy can be estimated to be around 39\,meV. One possible origin of such positive feature could be the kinematic exciton-exciton scattering mentioned above. Second, unlike the real and imaginary part of the 1Q nonrephasing spectrum in Figure~\ref{fig5}a and b, which show distinct absorptive and dispersive features, respectively, the real and imaginary part of the 2Q nonrephasing spectrum show mixed features. Such features are previously observed in gallium-arsenide quantum wells, which are ascribed to many-body interactions.\cite{karaiskaj2010two} Thirdly, a small side peak $\sigma$ at (1.736, 3.312)\,eV, is observed in the absolute diagram, while the real and imaginary part of the spectra show stronger signals. As the $\sigma$ peak absorbs approximately twice the $|e_1\rangle$ energy and emits at the $|e_2\rangle$, it suggests that the coherence originates between the $|e_1\rangle$ exciton complexes (i.e. unbound exciton pair $2|e_1\rangle$ or the bound exciton, $|f_1\rangle$) and the single exciton, $|e_2\rangle$. In contrast, we did not observe the coherences between the $|e_2\rangle$ exciton complexes and $|e_1\rangle$, although a slight elongation on top of the $\beta$ peak in Figure~\ref{fig5}e suggests its weak presence. Although the 2Q nonrephasing spectra provided rich information in the multi-exciton correlations, certain concerns still remain. One of which is the non-negligible spectral overlap between $\beta$ and $\sigma$ in Figure~\ref{fig5}d and e, leading to ambiguities in deciphering the many-body effects on their presence. 

It is worth mentioning that the biexciton states observed here do not originate from the higher-lying excited state. Previous transient absorption measurements on N2200 show that the excited-state absorption lies around 400\,meV above the ground-state absorption, which is outside the spectral window here.\cite{jin2017thermal} In addition, Denti \textit{et al.} previously conducted Raman and infrared spectroscopy on the doped N2200 system, showing that the polaron formation is strongly localized on the NDI units, in great contrast to other conjugated homopolymers.\cite{denti2019polaron} Despite the fact that the 2D 1Q measurements performed here look at exciton dynamics at initial population time ($T=20$\,fs), it is not unreasonable to hypothesize that the biexcitons observed in this work might be attributed to interactions between localized excitons on stacked NDI unit and its neighboring unit. Although we only probed one sample under specific processing conditions, further studies incorporating samples processed under different conditions will be valuable to correlate exciton dynamics with solid-state microstructure which will be essential to understand multi-exciton properties in semiconductor polymers. Previously, the short- and long-range aggregation in N2200 have been demonstrated to be tuned by varied molecular weights\cite{nahid2017unconventional}, solvent quality\cite{nahid2018nature}, film annealing\cite{trefz2018tuning}, blending\cite{tang2021morphology} and etc, which give handles to observe exciton pair and biexciton generations by different preparation processes. Furthermore, to assign the satellite vibrational mode, $\alpha$, both the energy separation between $\alpha$ and $\beta$ and Huang-Rhys (HR) factors are needed. Figure~\ref{fig5} suggests that the energy difference between the two vibronic excitons should be larger than 72\,meV (580\,cm$^{-1}$) as the side peak is limited by the spectral window. The accurate assignment of the vibrational mode other than the dominant ring-stretching mode is still elusive due to their unknown HR factors. Nonetheless, if we assume the HR factors of both modes are comparable, the first vibrational mode might land below 1000\,cm$^{-1}$, which is in wavenumber range of the low-energy stretching and torsional modes of the chain backbone, since the dominant Raman modes are already around 1500\,cm$^{-1}$ in N2200 thin film. Last but not the least, the mixed biexciton state does not seems to contribute significantly in either 1Q or 2Q spectra. As shown by Yang and Mukamel, spectral features from both mixed biexciton states should reside off-diagonally with equal two-quantum energy.\cite{yang2008revealing} However, the two biexcitons observed in this work do not show coherences from a mixed biexciton state. Only a small interstate coherence peak from pure $|e_1\rangle$ exciton complexes and the single exciton $|e_2\rangle$ is observed. Although the electronic transition from $|e_1\rangle$ or $|e_2\rangle$ to $|f_m\rangle$ is allowed, the vibrational transition from $|v_1\rangle$ to $|v_2,v_1\rangle$ could be partly forbidden due to the orthogonality of the two normal vibrational modes as indicated in Equation~\ref{eqn1}, where the equality holds true under Born-Oppenheimer (BO) approximation, leading to the weak and even no appearance of the coherences from the mixed biexciton. 
\begin{equation}\label{eqn1}
        \langle f_m; v_1, v_2|\vv{\mu}|e_1; v_1\rangle=\langle f_m|\vv{\mu}|e_1\rangle\langle v_1, v_2|v_1\rangle
\end{equation}
Previous work by De Sio \textit{et al} has demonstrated the presence of conical intersections of multiple potential wells addressed by both symmetric and asymmetric vibrational modes in molecular aggregates.\cite{de2021intermolecular} Close to the conical intersection does the BO approximation break down since the non-adiabatic transition is enabled by the vibronic coupling. However, as all measurements performed here are at early population time, the BO approximation should still hold considering that the coherent exciton motion does not initiate yet. Nevertheless, the 2Q spectral features at long population times are of great interest to investigate, as the conical intersection will allow transitions to dark states which are not visible under direct optical excitation. 

Finally, we highlight 
that the pump fluences employed in this work range from 10-100\,\fluenceunit, in which sufficient exciton-exciton annihilation (EEA) is expected in electron push-pull polymers on picoseconds time scale.\cite{zheng2023exciton, wang2021intrachain} Our work shows direct evidence of both correlated exciton pairs and bounded biexcitons even at initial population time, which might be precursors for EEA process in N2200. Dost\'al \textit{et al.} directly monitored the change of two-quantum peak intensities for a molecular aggregate in five-wave-mixing experiments with time evolving into the nanosecond range.\cite{dostal2018direct} By fitting the temporal evolution with the derived theoretical result considering the direct population of biexciton states, they were able to acquire an associated diffusion constant in good consistency with previous literature.

In addition to the method of direct monitoring through EEI, we suggest that the lineshape at initial population time and the diffusion constant might have a deterministic correlation. Moix \textit{et al.} studied the quantum transport behavior theoretically at short and long times in a one-dimensional J-aggregate chain, when both static disorder and environmental fluctuations exist. Of particular relevance, they treated either analytical solutions for master equations for the exciton dynamics, which correlate the exciton diffusion constants to the Coulombic coupling constant, static disorder and dephasing rates. In conjugated polymers, the first two parameters can theoretically be acquired by fitting the linear absorption spectra the with the Spano model.\cite{clark2009determining} Meanwhile, the dephasing rates could be determined by analyzing the full coherent line shape properly in 2DCS measurements by utilizing a microscopic theory of dephasing. 

The microscopic dephasing theory points out that the exciton dynamics generated by the impulsive excitation are not only determined by population decay, which are in turn determined by the radiative and nonradiative rates, but that there is also a contribution to decoherence due to system-bath interactions e.g.\ exciton-phonon and exciton-exciton scattering.\cite{moody2015intrinsic,li2023optical} The combination of both gives rise to the homogeneous linewidth in frequency domain, which can be determined by fitting the antidiagonal cut with a Lorentzian function in a purely homogeneously broadened limit. However, in addition to the homogeneous line broadening contributions, the inhomogeneous broadening arising from static disorder (e.g.\ each molecular segments adopts a slightly different conformation, resulting in different transition energies) will broaden the diagonal line shape, which has an impact on the antidiagonal linewidth concurrently.\cite{siemens2010resonance} Therefore, alongside the Coulomb coupling constant and the static disorder, the remaining parameter, homogeneous dephasing rate, could be obtained through the lineshape analysis, thus, an effective diffusion constant can be determined. The comparison between this and the one determined from traditional ultrafast measurements could lead to new physical insights into the evolution of exciton transport and diffusion behavior.


In conclusion, we perform 1Q and 2Q coherent optical spectroscopic measurements on an electron push-pull conjugated polymer, where clear features originating from two heterogeneous vibronic excitons alongside their exciton complexes are observed. 1Q measurements display spectral features due to the advantageous attractive bound biexcitons, leading to asymmetric cross peaks. The resultant 2D spectra can be explained qualitatively by tracing the Liouville pathways using a two-exciton model. The 2Q nonrephasing diagram provides further unambiguous evidence on both bound biexcitons and unbound exciton pairs. Specifically, unbound exciton pairs are found to be the dominant feature with a strong attractive biexciton subpeak, the binding energy of which is approximately 70\,meV. A weak repulsive biexciton is also observed from the shoulder of the unbound exciton pairs. The unbound exciton pairs show mixed absorptive and dispersive lineshape in contrast to that of the attractive biexciton, indicating the many-body effects in the unbound but correlated exciton pairs. 


\clearpage

\begin{acknowledgement}

CSA acknowledges funding from the Government of Canada (Canada Excellence Research Chair CERC-2022-00055) and from the Courtois Institute, Facult\'e des arts et des sciences, Universit\'e de Montr\'eal (Chaire de Recherche de l'Institut Courtois) for support during the redaction of the manuscript. CSA, ER, and ERG acknowledge support from National Science Foundation (NSF) Grant number DMR-2019444 (IMOD) for support for the data acquisition and analysis; YZ and ML acknowledge NSF-DMREF funding through grant number 1922111 for support for sample preparation and characterization. ER also appreciates support associated with the Carl Robert Anderson Chair funds at Lehigh University. 

\end{acknowledgement}

\begin{suppinfo}

See the Supplementary Information for the experimental methods including sample preparation, the technique of 2DCS and the XFROG results. Additional fluence-dependent 1Q and 2Q measurements are also provided for comparison.

\end{suppinfo}

\clearpage
\providecommand{\latin}[1]{#1}
\makeatletter
\providecommand{\doi}
  {\begingroup\let\do\@makeother\dospecials
  \catcode`\{=1 \catcode`\}=2 \doi@aux}
\providecommand{\doi@aux}[1]{\endgroup\texttt{#1}}
\makeatother
\providecommand*\mcitethebibliography{\thebibliography}
\csname @ifundefined\endcsname{endmcitethebibliography}
  {\let\endmcitethebibliography\endthebibliography}{}

\newpage


\section*{Supplementary material}

# Supporting Information: Unveiling Multi-Quantum Excitonic Correlations in Push-Pull Polymer Semiconductors

Yulong Zheng,[†] Esteban Rojas-Gatjens,[†] Myeongyeon Lee,[‡] Elsa Reichmanis,[‡] and Carlos Silva-Acuña[*,¶,†]

[†]*School of Chemistry and Biochemistry, Georgia Institute of Technology, 901 Atlantic Drive, Atlanta GA 30332, United States*
[‡]*Department of Chemical & Biomolecular Engineering, Lehigh University, 111 Research Drive, Bethlehem PA 18015, United States*
[¶]*Institut Courtois & Département de physique, Université de Montréal, 1375 Avenue Thérèse-Lavoie-Roux, Montréal, Québec H2V 0B3, Canada*

E-mail: carlos.silva@umontreal.ca

# Contents

# 1 Experimental Methods

## 1.1 Sample Preparation

N2200 ($M_w$ = 202261 g/mol, PDI = 2.22) was purchased from Ossila Limited. For sample preparation, a solution of 20 g/L in chlorobenzene (anhydrous, Sigma-Aldrich) was prepared by heating at 110°C for 30 minutes, followed by overnight aging at room temperature. The N2200 thin film was fabricated by blade coating on slide glass at 45°C, waiting until the solvent was fully evaporated. A shearing speed of 4mm/s was used to deposit the film.

## 1.2 Coherent Two-Dimensional Optical Spectroscopy

We employed the experimental setup designed by Turner *et al.*[1] The fundamental beam of 1030 nm was generated using an ultrafast laser system (Pharos Model PH1-20-02-10, Light Conversion) with a repetition rate of 100 kHz. Thereafter, the laser pulse centered around 710 nm was generated through a home-built second harmonic non-collinear optical parametric amplifier. The generated pulse was then directed to the diffractive optical element (DOE). With the zeroth-order beam blocked, the four first-order beams were residing on the four apexes of a square. The superior phase stability and control are achieved through the optical modulation by a reflective two-dimensional liquid crystalline spatial light modulator (SLM).[1] The four beams were compressed individually through the chirp scan[2] and multiphoton intrapulse interference phase scan.[3] Their temporal full widths at half maximum (FWHM) are then characterized through the second-harmonic cross frequency-resolved optical gating (XFROG) .[4] All samples are measured under a high-vacuum vibration-free cryostation at ambient temperature. (Montana Instruments)

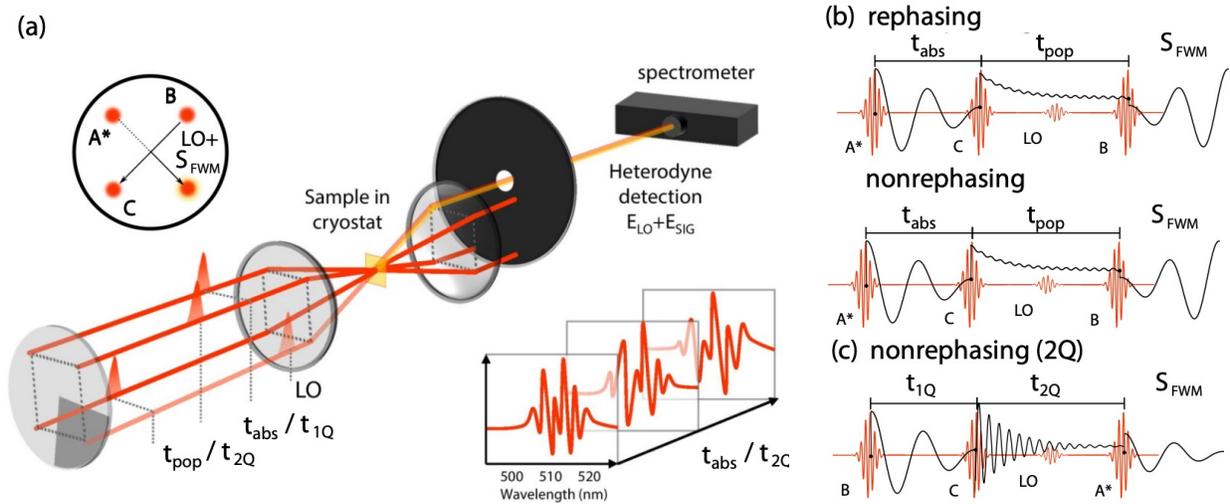

**Figure S1:** Schematic of the central part of the coherent optical beam recombination technique (COLBERT). (a) The four beams are propagating in the four corners of a square in parallel. Such geometry is called BOXCAR (Box Coherent Anti-Stokes Raman Scattering). The temporal delays, $t_{pop}$ and $t_{abs}$, and $t_{2Q}$ and $t_{1Q}$ between beams are controlled by the SLM. With such phase relation, the four-wave-mixing signal, $S_{FWM}$ is emitted in the wavevector-conserved direction of the first three beams. The signal is acquired through the heterodyne-detection by interfering with the local oscillator, (LO). (b) The pulse sequences for 1Q rephasing and nonrephasing direction, respectively. (c) Using the same geometry, the 2Q nonrephasing signal can be acquired with alternating pulse sequence. The asterisk indicates the phase-conjugated beam (A*). Figure adapted from Ref. 5 with permission.

## 1.3 Pulse Width Characterization

The temporal full widths at half maximum for the compressed pulses are fitted by a Gaussian fit, found to be around 17 fs, characterized by XFROG. (Figure S2).

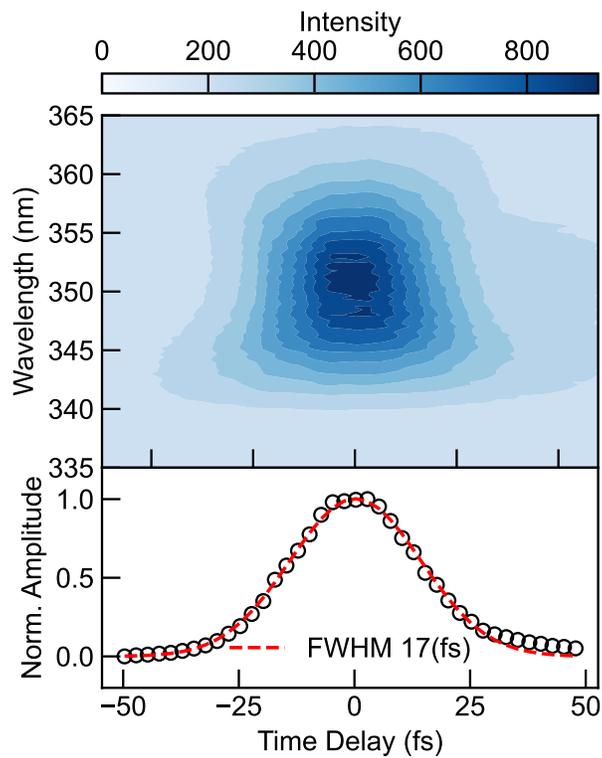

**Figure S2:** The compressed pulse estimated by a Gaussian fit, characterized by XFROG.

# 2 Additional Experimental Results

## 2.1 Fluence-Dependent 1Q Rephasing

The complementary set of fluence-dependent 1Q rephasing diagrams is shown in Figure S3. To more clearly show the dependence of the spectral linewidths, we take the antidiagonal and diagonal cuts of the absolute values of the 1Q rephasing diagrams as displayed in Figure S4. For the antidiagonal cuts, we can readily observe no drastic variance in the spectral linewidths with varying fluences.

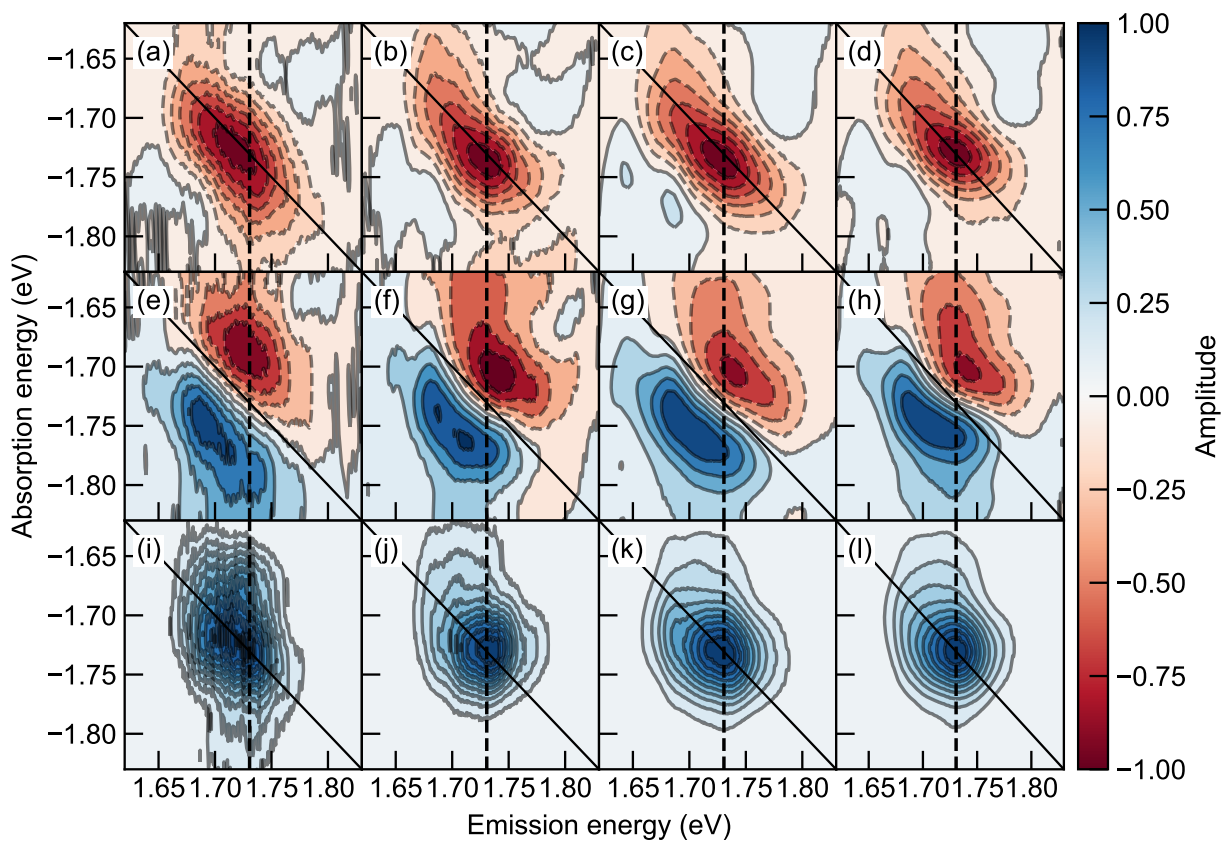

**Figure S3:** Fluence-dependent 1Q rephasing diagrams. The three rows are real (a-d), imaginary (e-h) and absolute (i-l) values of the 2D spectra. The four columns measured under varying fluences. From left to right are 6.4, 12.7, 51.2 and $121.0\,\mu J/cm^2$.

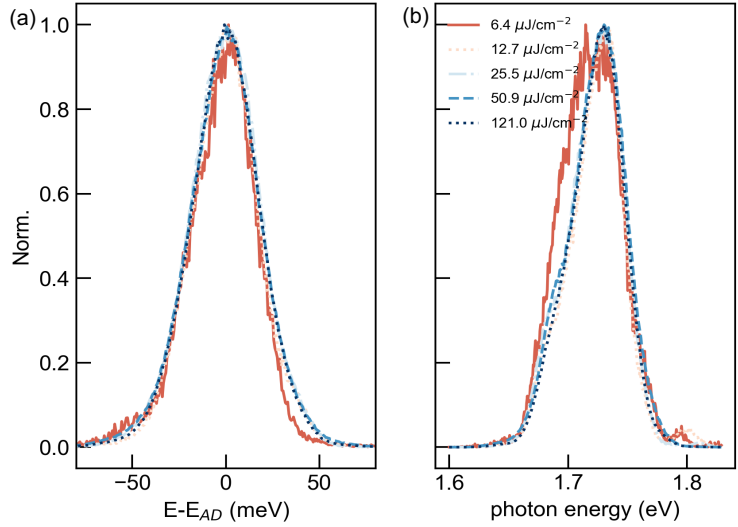

**Figure S4:** Fluence-dependent antidiagonal (a) and diagonal cuts (b) of the absolute-valued diagrams as shown in Figure S3.

## 2.2 Fluence-Dependent 2Q Nonrephasing

Here, we show the 2Q measurements performed from low to high fluences (12.7, 25.6, and 51.2 and 121 $\mu J/cm^2$) as shown in Figure S5, the last of which is the same as shown in Figure 5 in the main article, displaying here for a better comparison. All measurements share the dominant features of the two diagonal peaks indicating the two heterogeneous vibronic peaks, each associated with a red-shifted side peak indicating the existence of attractive biexciton states. In addition, certain features are evolving with increasing fluences, including the blue-shifted side peak around (1.66, 3.37)eV and off-diagonal feature at (1.73, 3.30)eV. There could be a possibility of them being an artefact at low fluences, as the modulation efficiency of the spatial light modulator depends on the intensity of the incident light beam.[6] There could also exist possibility of unknown physical processes, which mediate the coupling of the two excitons by fluences. Further quantum dynamics simulations and high-order-wave-mixing experiments are needed to characterize this process.

To show the correlated exciton pairs, attractive and repulsive biexcitons more clearly, we take the vertical cut of the Figure S5d at $E_{1Q}$=1.736 eV and 1.666 eV. The x-axis is directly adjusted by relation $E_{2Q} - 2E_{1Q}$. Take the dominant vibronic exciton pairs ($E_{1Q}$=1.736 eV

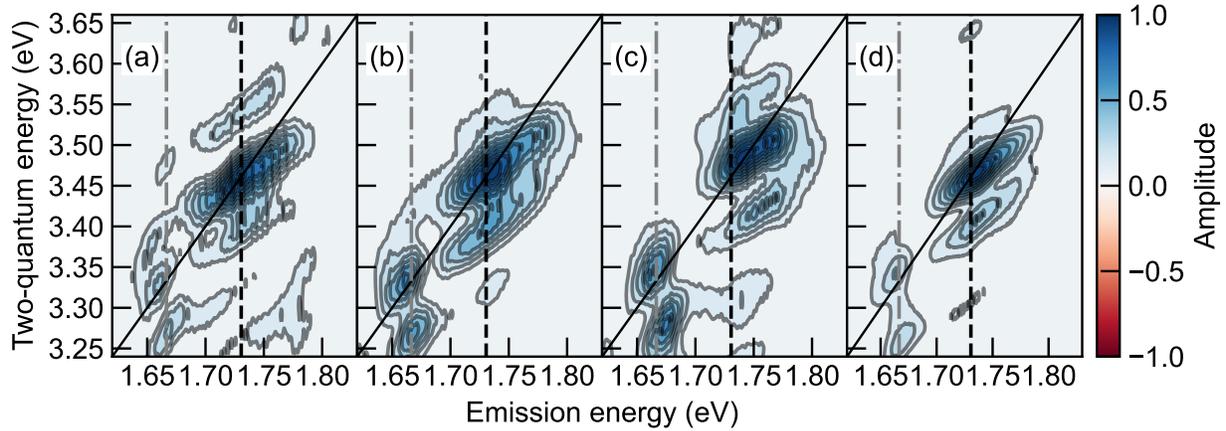

**Figure S5:** Fluence-dependent 2Q absolute-valued spectra measured under (a) 12.7, (b) 25.6, (c) 51.2 and (d) 121 $\mu$J/cm$^2$, respectively. The black dashed line indicate the peak position at the dominant $A_{0-1}$. The grey dashed-dotted line locates at the side peak position.

for example, it can be clearly seen that beside the dominant exciton pairs at $\Delta E = 0$ eV, which is the exact double of the $E_{1Q}$, a red-shifted peak is observed at around -64 meV, indicating an attractive biexciton. In addition, a blue-shifted side peak at 39 meV can also be observed, suggesting the repulsive biexciton state. The side peaks could be observed more clearly in Figure S5a-c, probably with more noise, nevertheless.

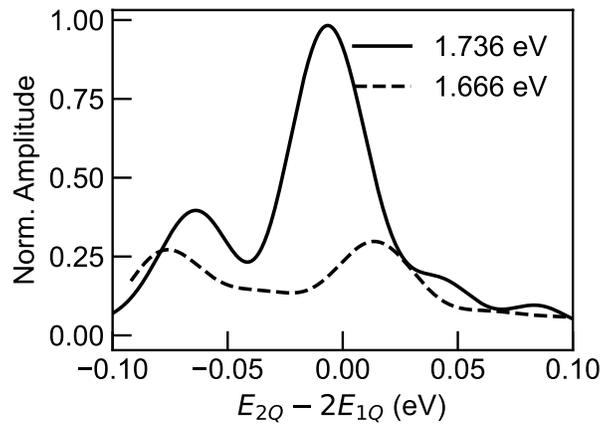

**Figure S6:** Vertical cuts at $E_{1Q}$=1.736 eV (solid) and 1.666 eV (dashed) of Figure S5d for the dominant vibronic exciton pairs associated with the both attractive and repulsive biexciton states.



\end{document}